\documentclass[10pt,a4paper]{article}
\usepackage{jheppub}

\usepackage{epsfig}
\usepackage{latexsym}
\usepackage{amsfonts}
\usepackage{amsmath}
\usepackage{amsthm}
\usepackage{amssymb}
\usepackage{amsbsy}
\usepackage{multirow}
\usepackage{slashed}
\usepackage{amsmath,latexsym,amssymb}
\usepackage{graphicx}
\usepackage{psfrag}
\usepackage[latin1]{inputenc}
\usepackage{nicefrac}
\usepackage[stable]{footmisc}
\usepackage{braket}
\usepackage{rotating}
\usepackage{afterpage}
\usepackage{bbm}

\def\be{\begin{equation}}
\def\ee{\end{equation}}
\def\bea{\begin{eqnarray}}
\def\eea{\end{eqnarray}}
\def\ba{\begin{eqnarray}}
\def\ea{\end{eqnarray}}

\newcommand{\ft}[2]{{\textstyle\frac{#1}{#2}}}

\def\be{\begin{equation}}
\def\ee{\end{equation}}
\def\bea{\begin{eqnarray}}
\def\eea{\end{eqnarray}}
\def\ba{\begin{array}}
\def\ea{\end{array}}
\def\bd{\begin{displaymath}}
\def\ed{\end{displaymath}}

\def\e{\epsilon}           

\def\g{\gamma}

\def\l{\lambda}

\def\m{\mu}
\def\n{\nu}
  
\def\pa{\partial}                              

\def\>{\rangle} 
\def\<{\langle} 
\def\Dsl{D \hskip-.6em \raise1pt\hbox{$ / $ } }

\def\pa{\partial}

\newsavebox{\uuunit}
\sbox{\uuunit}
    {\setlength{\unitlength}{0.825em}
     \begin{picture}(0.6,0.7)
        \thinlines
        \put(0,0){\line(1,0){0.5}}
        \put(0.15,0){\line(0,1){0.7}}
        \put(0.35,0){\line(0,1){0.8}}
       \multiput(0.3,0.8)(-0.04,-0.02){12}{\rule{0.5pt}{0.5pt}}
     \end {picture}}

\def\be{\begin{equation}}
\def\ee{\end{equation}}
\def\bea{\begin{eqnarray}}
\def\eea{\end{eqnarray}}

\newcommand{\beq}{\begin{eqnarray}}
\newcommand{\eeq}{\end{eqnarray}}

\def\g{\gamma}

\def\e{\epsilon}

\def\l{\lambda}

\def\m{\mu}
\def\n{\nu}

\def\E {$E_{7(7)}$}

\def\pa{\partial}

\def\sF{{{ F}\!\!\!\!\hskip.8pt\hbox{\raise1pt\hbox{/}}\,}}
\def\som{{{ \omega}\!\!\!\!\hskip.8pt\hbox{\raise1pt\hbox{/}}\,}}
\def\sJ{{{\rm J}\!\!\!\!\hskip.8pt\hbox{\raise1pt\hbox{/}}\,}}

\newcommand{\Nt}{{\ensuremath{\mathcal N{=}2}\ }}

\title{{ {\textsc{N=2 Supergravity Counterterms,  Off and On Shell }} }}
\author[a,b]{W.  Chemissany,}  \author[c,d]{S. Ferrara,} \author[a]{R.  Kallosh,} \author[a,f]{C. S. Shahbazi}
\affiliation[a]{Stanford Institute for Theoretical Physics and Department of Physics, Stanford University,\\ Stanford, CA 94305-4060, USA} 
\affiliation[b]{Department of Physics and Astronomy, University of Waterloo, Waterloo, Ontario, Canada, N2L 3G1}
\affiliation[c]{Physics Department, Theory Unit, CERN, CERN, CH 1211, Geneva 23, Switzerland} 
\affiliation[d]{INFN - Laboratori Nazionali di Frascati, Via Enrico Fermi 40, I-00044 Frascati, Italy}
\affiliation[f]{Instituto de Fisica Teorica UAM/CSIC, C.U.~Cantoblanco, 28049 Madrid, Spain}

\emailAdd{chemissany.wissam@gmail.ca} \emailAdd{sergio.ferrara@cern.ch} \emailAdd{kallosh@stanford.edu}\emailAdd{carlos.shabazi@uam.es}

\abstract{
 We study N=2 supergravity deformed by a genuine supersymmetric completion of the $\lambda R^4$  term, using  the underlying off shell  N=2 superconformal framework.  The gauge-fixed superconformal model has unbroken local supersymmetry of N=2 supergravity with higher derivative deformation. Elimination of  auxiliary fields leads to the deformation  of the supersymmetry rules as well as to the deformation of the action, which becomes  a Born-Infeld with higher derivative type action. We find that the gravitino supersymmetry deformation starts from 
 $\lambda \, \pa^4 {\cal F}^3$ and has higher graviphoton couplings. In the action there are terms $\lambda^2 \pa^8  {\cal F}^{6}$ and higher, in addition to original on shell counterterm deformation.  These deformations are absent  in the on shell superspace and in the candidate on shell counterterms of N=4,~8  supergravities, truncated down to N=2. We conclude therefore that the undeformed on shell superspace candidate counterterms  break the N=2 part of local supersymmetry.  }

\begin{document}
 \maketitle
\section{Introduction}

The notorious $(R_{....})^4$ counterterm in supergravity had its ups and downs since the time it was first proposed as  a candidate for the UV divergence in N=1 supergravity \cite{Deser:1977nt} back in 1977.  For N=2 a linearized version of the candidate for the UV divergence was proposed a year later in \cite{Deser:1978br}. The linearized version of it for N=8 supergravity was  constructed in \cite{Kallosh:1980fi,Howe:1981xy}. The gravitational part of it is the square of the Bel-Robinson tensor and it also has a term quartic in graviphotons of the form $[\pa {\cal F}]^4$. In N=8 supergravity  the 3-loop UV finiteness was established a while ago in \cite{Bern:2007hh,Bern:2008pv} and was explained via \E duality. Two type of explanations were given. One was based on the fact that the 3-loop counterterm breaks \E duality at the non-linear level \cite{Brodel:2009hu,Elvang:2010kc,Beisert:2010jx},  the other was based on the fact that the 3-loop counterterm breaks the Noether-Gaillard-Zumino \E\,  deformed duality current 
conservation \cite {Kallosh:2011dp}. 

Fully non-linear candidate counterterms starting from L=N number of loops were constructed in \cite{Kallosh:1980fi,Howe:1980th} based on the Lorentz covariant on shell\footnote{Since the topic of this paper  has to do with ``off shell and on shell counterterms'', we will give a clear definition of these two concepts in sec. 2 where  the subtleties of these definitions will be explained when the action is deformed by the higher derivative terms and when the auxiliary fields are available. In short, the on shell N-extended superspace in \cite{Brink:1979nt,Howe:1981gz} has no auxiliary fields and the physical fields satisfy classical undeformed EOM (equations of motion).}
N-extended superspace geometry \cite{Brink:1979nt,Howe:1981gz}.
The complete non-linear version of the candidate $R^4$ counterterm in N=4 supergravity was constructed only recently in \cite{Bossard:2011tq}, just in time to be ruled out half a year later as a UV divergence of the 3-loop $N=4$ supergravity  \cite{Bern:2012cd,Tourkine:2012ip}.

One possible explanation of the 3-loop N=4 UV finiteness in \cite{Kallosh:2012ei} is via the  $SL(2, \mathbb{R})\times SO(6)$ duality symmetry, broken by the presence of the counterterm, which violates the Noether-Gaillard-Zumino current conservation. However in \cite{Bossard:2011ij} it was suggested that one can restore duality invariance of the full theory by adding an infinite number of additional terms to the theory. This implies that instead of the original version of the theory one should construct its Born-Infeld type generalization \cite{Carrasco:2011jv,Chemissany:2011yv,Broedel:2012gf}. While this generalization can indeed be constructed for  N = 2, as we will show in this paper,  it is not known how to generalize this restoration of duality for N = 4, 8. It was also shown in \cite{Kallosh:2012yy} that it is impossible to restore duality symmetry, broken by the counterterm, while preserving the framework of the on shell superspace \cite{Brink:1979nt,Howe:1981gz}.

Nevertheless, one could argue that, unlike local gauge symmetry, global duality symmetry may or may not be preserved at the quantum level even if there are no anomalies. Therefore it would be important to have an independent explanation of the absence of the  $R^4$ counterterm in N=4 supergravity, which may be formulated using more familiar tools based on local  supersymmetry.

In this paper we will show that the same type of arguments as the one used in \cite {Kallosh:2011dp,Kallosh:2012ei} with respect to global duality, can be formulated in extended supergravities by imposing the condition of local supersymmetry. For definiteness, we will describe the new construction in terms of N = 2 supergravity, where the off shell theory is available. The main argument will apply to all extended supergravities truncated to N=2 supersymmetry. We will show that whereas the  on shell counterterm candidates constructed in the earlier works  are invariant with respect to the classical undeformed local supersymmetry transformation under condition that classical EOM are satisfied, the classical theory complemented by such counterterms is no longer supersymmetric. 

We will find that the  genuine\footnote{We define the ``genuine'' supersymmetric action as the action which is invariant under local symmetries, which requires that the fields of the theory do not satisfy their EOM. There are two versions of such ``genuine'' actions: the first one  involves independent off shell  auxiliary fields, see eq.  (\ref{susyDef}), the second one has auxiliary fields satisfying their EOM, see eq.  (\ref{susyD}). In both cases the physical, not auxiliary fields, should not satisfy any equation, when the variation of the action is computed. Meanwhile, the on shell counterterms have a property that the physical fields satisfy the classical EOM, therefore {\it a priory} they are not ``genuine''. } supersymmetry forbids the known on shell counterterms, unless one can formulate a consistent supersymmetric Born-Infeld generalization of the initial classical theory. This is similar to duality arguments, but now instead of duality, we may use a much better understood requirement of local supersymmetry. 
This may help to understand why the early analysis based on the Lorentz covariant on shell superspace is not sufficient to explain the actual amplitude computations \cite{Bern:2007hh,Bern:2008pv,Bern:2012cd,Tourkine:2012ip}.

In extended supergravities at N $\geq 3$ only the on shell Lorentz covariant superspace is known, which means that the Bianchi identities were solved  under condition that classical EOM  are valid, both in the linearized approximation as well as in the full non-linear case. Therefore, if we would encounter a UV divergence associated with the on shell Lorentz covariant superspace counterterm,   we would have to add such a term to the classical action to be able to absorb this UV divergence.

It is not clear if the corresponding classical supergravity action deformed by a higher derivative supersymmetric invariant actually possess an N-extended local Q-supersymmetry since the higher derivative action is supersymmetric only when the classical EOM are valid  at N $\geq 3$. Meanwhile, all local symmetries have to be valid off shell, therefore the known form of the candidate counterterms by no means guarantees that the classical plus higher derivative extended supergravity action has a local supersymmetry\footnote{In pure gravity the  counterterm $(R_{....})^3$ is locally general covariant irrespectively as to whether  the on shell condition $R_{\mu\nu}= R=0$ is used or not. Therefore in pure gravity the  classical action deformed by the ``on shell counterterm''  $(R_{....})^3$  is general covariant. But this is not the case for the extended classical supergravities deformed by the higher derivative supersymmetric invariant, since the known counterterms are locally supersymmetric only when classical 
EOM are valid.}.

The purpose of this paper is to find out if the classical N-extended supergravity action deformed  by the on shell candidate  counterterm has an unbroken N=2 part of the local supersymmetry. To explain the issue,  we split all fields 
\be
\phi = \{ \phi_{phy}\, ,  \phi_{aux} \}
\ee
into physical $\phi_{phy}$ and auxiliary $\phi_{aux}$, the last ones  are non-propagating at the classical level, but support the off shell local N=2 supersymmetry \cite{Fradkin:1979cw}.

We will show, by comparing the action with the genuine N=2 supergravity with higher derivative $(R_{....})^4$ counterterm that the action with the ``on shell counterterm deformation''
\be
S_1^{def} (\phi_{phy})= S_{0}(\phi_{phy}) + \lambda S_{ct}(\phi_{phy})
\label{onc}\ee
is not supersymmetric under classical ``on shell'' supersymmetry transformations $\delta_Q^{cl}$
\be
\delta^{(0)}_Q  S_1^{def} (\phi_{phy})\neq 0.
\ee
The supersymmetric one requires  $\lambda$ deformation of the classical action
\be
S^{def} (\phi_{phy})= S_{0}(\phi_{phy}) + \lambda S_{ct}(\phi_{phy}) +\sum_{n=2}\lambda^n S^{(n)}(\phi_{phy})
\label{Sdeform},\ee
simultaneously with  $\lambda$ deformation of the gravitino supersymmetry transformation
\be
\delta_Q^{def} \psi_\mu ( \phi_{phy})= \delta_Q^{(0)}\psi_\mu(\phi_{phy})  + \sum_{n=1} \lambda^n \delta_Q^{(n)}\psi_\mu(\phi_{phy})
\label{susydef}.\ee
In this paper  we will identify in N=2 supergravity with $\lambda R^4$ deformation the following new terms  required for local supersymmetry
\be
 \lambda^2 S^{(2)}(\phi_{phy}) \, ,  \qquad \lambda \delta_Q^{(1)}\psi_\mu(\phi_{phy}).
\label{new}\ee
We will provide an explicit non-vanishing expressions for these terms which are not available in the on shell superspace construction \cite{Brink:1979nt,Howe:1981gz} and in the corresponding on shell counterterms \cite{Kallosh:1980fi,Howe:1980th,Bossard:2011tq}.

In N=2 such a consistent all order in $\lambda$ deformation reconstructing the supersymmetry broken by the on shell counterterm is possible since the genuine supersymmetric action and susy rules  are known in the closed form, before the auxiliary fields are replaced by their $\lambda$-dependent values. In N=2 supergravity this deformation procedure is producing a Born-Infeld type supergravity with higher derivatives and all powers of $\lambda$ and graviphotons, as we will explain below, once the supersymmetric $\lambda R^4$ term is added to the N=2 supergravity action.

For N=4 and N=8 a consistent deformation reconstructing the supersymmetry broken by the on shell counterterm is a challenge. Our work here will  prove that without an additional deformation of the on shell superspace and the on shell candidate counterterms  supersymmetry of the classical action with counterterm is broken. The analogous argument from the global duality symmetry was given in \cite {Kallosh:2011dp}, however, local supersymmetry is much more familiar concept and therefore the arguments that the currently known ``on shell superspace candidate counterterms'' break supersymmetry should be easier to confirm/disprove.

\section{Auxiliary fields and deformation}
In N=2 case the genuine
superconformal action with higher derivative deformation is known, see for example the superspace construction of $R^4$  in \cite{Moura:2002ip,Kuzenko:2008ry,Butter:2010jm}  and the superconformal one in \cite{deWit:2010za}.  In both cases the theory is not yet given in the form shown in (\ref{Sdeform}), (\ref{susydef}) since in both cases the actions depend on auxiliary fields.

We may write the genuine N=2 supersymmetric action with higher derivatives as
\be
S^{def}(\phi_{phy}, \phi_{aux})= S^{(0)} (\phi_{phy}, \phi_{aux}) +\lambda S^{(1)} (\phi_{phy}, \phi_{aux}),
\ee
and 
\be
\delta_Q^{def} \psi_\mu = \delta_Q\psi_\mu (\phi_{phy}, \phi_{aux}).
\label{susydefaux}\ee
One may solve EOM for auxiliary fields recursively, so that
\be
{\delta S^{def}\over \delta\phi_{aux}}=0\, , \qquad \Rightarrow \qquad \hat \phi_{aux}(\phi_{phy}, \lambda)= \phi_{aux}^{0}+ \sum_{n=1} \lambda^n
\phi_{aux}^{n}.
\ee
The action remains N=2 supersymmetric when the $ \phi_{aux}$ are replaced by their values $\hat \phi_{aux}(\phi_{phy},\lambda)$ solving the deformed EOM.
\be
S^{def}\Big (\phi_{phy}, \hat \phi_{aux}(\phi_{phy}, \lambda)\Big )= S_{0} \Big (\phi_{phy}, \hat \phi_{aux}(\phi_{phy}, \lambda)\Big) +\lambda S^4 \Big (\phi_{phy}, \hat \phi_{aux}(\phi_{phy},\lambda)\Big ),
\ee
or
\be
S^{def}\Big (\phi_{phy}, \hat \phi_{aux}(\phi_{phy}, \lambda)\Big )= S_{0}  (\phi_{phy}) +  \lambda S_{ct} (\phi_{phy}) + \lambda^2 S_2 (\phi_{phy})+\cdots,
\label{Sdeform1}\ee
and 
\be
\delta_Q^{def} \psi_\mu \Big (\phi_{phy}, \hat \phi_{aux}(\phi_{phy}, \lambda)\Big ) = \delta_Q^0\psi_\mu (\phi_{phy}) +
\lambda \delta_Q^1\psi_\mu (\phi_{phy})+\cdots.
\label{susydefaux}\ee
The deformed action depending only on $\phi_{phy}$ acquires a form (\ref{Sdeform}), (\ref{Sdeform1}) with  higher powers of $\lambda$ terms. 

Local supersymmetry means that
\be
\delta_Q S^{def} = \int d^4 x {\delta S^{def}\over \delta \phi^i(x) } \delta_Q \phi^i(x)=\int d^4 x {\delta S^{def}\over \delta \phi^i_{phy}(x) } \delta_Q \phi^i_{phy}(x) + \int d^4 x {\delta S^{def}\over \delta \phi^i_{aux}(x) } \delta_Q \phi^i_{aux}(x)=
0
\label{susyDef}.\ee
 When auxiliary fields satisfy their deformed EOM,  ${\delta S^{def}\over \delta\phi_{aux}}=0$,  the condition of local supersymmetry is
\be
\delta_Q S^{def} = \int d^4 x {\delta S^{def}\over \delta \phi^i(x) } \delta_Q \phi^i(x)=\int d^4 x {\delta S^{def}\over \delta \phi^i_{phy}(x) } \delta_Q \phi^i_{phy}(x) =
0
\label{susyD}.\ee
Requirement of local symmetry requires that the physical fields are off shell:
\be
\delta_Q S^{def}=0\, ,\qquad {\delta S^{def}\over \delta \phi^i_{phys}(x) }\neq 0
\label{off},\ee
since the variation of the action upon elimination of auxiliary fields is proportional to the variation of the  deformed action over 
 physical fields.  
 
 On shell superspace  \cite{Brink:1979nt,Howe:1981gz} which is a basis for constructing the N-extended supergravity counterterms in \cite{Kallosh:1980fi,Howe:1980th,Howe:1981xy,Bossard:2011tq} describes only the physical states of the theory, satisfying classical EOM. This means in the context of N=2 on shell superspace  the auxiliary fields take the values satisfying their classical EOM
\be
{\delta S_{0}\over \delta\phi_{aux}}=0\, , \qquad \Rightarrow \qquad \phi_{aux}= \phi_{aux}^{0}.
\ee
  Therefore the on shell counterterms \cite{Kallosh:1980fi,Howe:1980th,Bossard:2011tq}
 depending on $\phi^i_{phys}$ satisfying the classical on shell condition
\be
{\delta S_0\over \delta \phi^i_{phys}(x) }= 0
\ee
do not necessarily provide an unbroken local supersymmetry of the deformed action, which requires that (\ref{off}) takes place.

The gravitino supersymmetry transformation depends on the auxiliary fields, therefore the same replacement of $ \phi_{aux}$  by the values $\hat \phi_{aux}(\phi_{phy},\lambda)$  will render the susy rules $\lambda$ dependent as shown in (\ref{susydef}).

Thus if we take a known on shell superspace counterterm in N=8 or N=4 extended supergravity and truncate it down to N=2 supersymmetry we will get, following the old superspace counterterm paradigm, that
\be
S_1^{def}(\phi_{phy})= S_{0} (\phi_{phy})+ \lambda S_{ct}(\phi_{phy})
\label{oldS},\ee
\be
\delta_Q\psi_\mu= \delta_Q^{0}\psi_\mu
\label{oldSusy},\ee
and there is no need and no place  for  $\lambda^n$  terms with $n\geq 2$  in the action and no need and no place for terms $\delta_Q^{n}\psi_\mu$  with $n\geq 1$ in gravitino susy rules: on shell superspace  \cite{Brink:1979nt,Howe:1981gz} and the counterterms \cite{Kallosh:1980fi,Howe:1980th,Howe:1981xy,Bossard:2011tq} are not deformed but still viewed as legitimate candidates for UV divergences.

Thus, if we will be able to find, for example,  terms in the genuine supersymmetric action of  order $\lambda^2$ and in the gravitino susy transformation of the order $\lambda$, which are present there and absent in on shell counterms, it would prove that the deformation of the action with the on shell counterterm without deformation of the supersymmetry rules and without an additional deformation of the action, breaks the N=2 part of the N-extended supersymmetry.
We will show that instead of (\ref{oldS}), (\ref {oldSusy}) the genuine supersymmetric theory with higher derivatives starting with $R^4$ requires  the  deviation from  the ``on shell superspace counterterm" paradigm as shown in (\ref{susydef}) and in (\ref{new}).

\section{Superconformal off shell N=2 symmetry as a tool}

 In N=2 supergravity the off shell supersymmetry as well as an underlying N=2 off shell local superconformal symmetries are known.  Moreover, recently a complete non-linear superconformal N=2 invariant $R^4$ action,  was explicitly presented in \cite{deWit:2010za}. Many other higher derivative superconformal invariants can be constructed using the procedure described in \cite{deWit:2010za}, for example, $R^6$, $R^8$ and others which are interesting in the context of pure N=2 supergravity. The useful description of the superconformal N=2 models can be found in \cite{Freedman:2012zz} and in \cite{Mohaupt:2000mj}.

 One could also use the off shell superspace construction of the N=2 $R^4$ invariant in \cite{Moura:2002ip,Kuzenko:2008ry,Butter:2010jm} which is expected to produce the analogous results. The main reason here to use the superconformal N=2 tensor calculus and the recent construction of $R^4$ action in  \cite{deWit:2010za} is that the component results, especially the Born-infeld type graviphoton part of the action, are relatively easy to extract.

Here we will use the superconformal construction to study the action of the minimal (for simplicity) pure N=2 supergravity
without matter multiplets and with the simplest duality group $U(1)$, deformed by the off shell supersymmetric higher derivative action, quartic in curvatures.   The superconformal action, associated with $SU(2,2|2)$  algebra has many local symmetries: local D-dilatation, local chiral $U(1)$ symmetry, local $SU(2)$ symmetry, local S-supersymmetry and  local 
K-conformal symmetry in addition to  Q-supersymmetry, general covariance and Lorentz symmetry of N=2 Poincar\'e supergravity. Local dilatation, local chiral symmetry, local $SU(2)$ symmetry, local S-supersymmetry and  local conformal boost have to be gauge-fixed so that only Q-supersymmetry, general covariance and Lorentz symmetry remain in N=2 Poincar\'e supergravity.

\subsection{Outline of the construction of the higher derivative supersymmetric Born-Infeld type N=2  supergravity}

A general class of superconformal higher derivative actions including $R^4$ terms is given in \cite{deWit:2010za} and it is rather complicated, in general. Therefore in order to achieve our purpose of constructing higher derivative supersymmetric Born-Infeld type N=2  supergravity we have to do three things.

1. We  take a simplest choice of the class of N=2 superconformal actions in \cite{deWit:2010za}. The holomorphic prepotential $F$ depends on a single conformal compensator $X$ and is given by
 \be
 F=- {i\over 4}  X^2,
 \label{prep}\ee
 so that after we gauge-fix the extra superconformal symmetries not present in supergravity we will be left with pure N=2 minimal supergravity. This model has a graviton, 2 gravitino's and a graviphoton in a pure N=2 supergravity multiplet.
 To embed this model into the superconformal framework and to be able to add to the classical action the higher derivative superconformal version of the $R^4$ we  make a choice of the  generalized K\"ahler potential  
 $\mathcal{H}$ 
  \cite{deWit:2010za} which is defined by the conformal multiplet  of the Weyl weight $w=0$. Our choice is  
  \begin{equation}\mathcal{H}=\frac{({\cal T}^{-})^2}{X^{2}}\frac{({\cal T}^{+})^{2}}{\bar{X}^{2}}
\label{H}.\end{equation}
 Here ${\cal T}$ is the auxiliary field of the Weyl multiplet $W$ and $X$ is the first component of the chiral compensator superfield $S$, see Appendix A where these multiplets are described.

2. We  gauge-fix  superconformal symmetries which are not the local symmetries of N=2 supergravity. The choice of the second compensator is a non-linear vector multiplet, which is convenient when higher derivative action is present. Another choice we are making is for special S-supersymmetry, which provides us with explicit expressions for the Poincar\'e supersymmetry as a combination of superconformal Q- and S-supersymmetry.

3. We eliminate the auxiliary fields on their deformed EOM. We find that it is actually a recursive procedure generating higher powers with derivatives of the graviphoton, which is the reason to call the resulting model a ``higher derivative supersymmetric Born-Infeld type N=2 supergravity''.

Once this program  is accomplished, we  know the properties of the genuine  supersymmetric completion of the $R^4$ action in N=2 Poincar\'e supergravity, and can see how different it is from the on shell one.

Our starting point is a   superconformally invariant \Nt  action  \begin{equation}
  \int d^4\theta \left(S^2+\lambda\frac{W^2}{S^2}\mathbb{T}\left(\frac{\overline{W^2}}{\overline{S^2}}\right)\right)\,.
  \label{faction}
\end{equation}
Here  the first term, the chiral multiplet action,  interacting with the Weyl multiplet, is the N=2 superconformal action which after gauge-fixing the extra symmetries becomes a Poincar\'e N=2 supergravity classical action. The second term is defined in  \cite{deWit:2010za} and is known as the kinetic chiral multiplet action. 
Our $w=0$ chiral multiplet $\Phi$ is the ratio of the square of the Weyl multiplet $W^2$ to the square of the chiral compensator multiplet $S^2$,
$\lambda$ is the deformation parameter. We provide details of this construction in the Appendix B.

To study more general superconformal higher derivative action described in  \cite{deWit:2010za},  one could either take for the $w=0$ chiral multiplet $\Phi$ some more general functions of the Weyl multiplet $W^2$ and of the chiral compensator multiplet $S^2$ in the kinetic chiral multiplet action there, or to use the more general class of actions for the
composite chiral and anti-chiral multiplets with suitable Weyl weights. However, the analysis of the counterterm which we perform for $R^4$, excluding auxiliary fields to compare with the on shell ones, has to be perfumed on more general superinvariants separately.

\subsection{Simple reason for Born-infeld type action with ${\cal T}$ eliminated on deformed EOM}
The superconformal  action (\ref{faction}) has terms like
\be
 X\bar X \, R \, , \qquad  \,  \lambda (X \bar X)^{-2} R^4 \, , \qquad  \, \lambda (X \bar X)^{-2} (\pa {\cal T}  \pa {\cal T})^2 \,,
\ee
When the local dilatation and local chiral symmetry are fixed by the requirement that
$
X=\bar X=1
$
in Planck units, these terms in the action become, respectively, proportional to 
\be
 R \, , \qquad \, \lambda  R^4 \, , \qquad \, \lambda  (\pa  {\cal T}  \pa{\cal T})^2  \,.
 \ee
The existence of the {\it term proportional to $\lambda$  quartic in} ${\cal T}$ is of crucial importance here. The explicit form can be deduced either from the superconformal action  (\ref{faction}), see Appendix B,  or can be looked up in \cite{Chemissany:2011yv} 
where one finds that it is proportional to 
\begin{eqnarray} \lambda \, [\pa {\cal T}]^4\equiv \lambda\,  t_{\mu_1 ... \nu_4}  &&[\pa_{\mu}{\cal T}^{+\,\mu_1 \nu_1}\pa^{\mu}{\cal T}^{-\, \mu_2 \nu_2}\pa_{\nu} {\cal T}^{+\,\mu_3 \nu_3}\pa^{\nu}{\cal T}^{-\, \mu_4\nu_4}\nonumber \\
\nonumber \\
&&+ \, \frac{1}{2}\, \pa_{\mu}{\cal T}^{+\,{\mu_1 \nu_1}} \pa^{\mu}{\cal T}^{+\,\mu_2 \nu_2}\pa_{\nu} {\cal T}^{-\,\mu_3 \nu_3}\pa^{\nu}{\cal T}^{-\, \mu_4\nu_4}]\label{CKO},\end{eqnarray}
where $t_{\mu_1 ... \nu_4}\equiv t_{\mu_1 \nu_1 \mu_2 \nu_2 \mu_3 \nu_3 \mu_4\nu_4}$ is the $t_8$-tensor antisymmetric in the pairs $\mu_i \nu_i$   and symmetric under
the exchange of such pairs.
 It has been checked in  \cite{Chemissany:2011yv} that this term in the approximation that ${\cal T}\sim {\cal F}$  is the same as the explicit quartic graviphoton part of the N=8 $R^4$ counterterm, truncated  to $U(1)$ from $SU(8)$. In particular, it defines the linearised quartic N=2 supersymmetric partner of the $\lambda R^4$ higher derivative term.

Every term in the deformation of the action has 4 powers of the Weyl multiplet. In terms of components it means that there are at least four powers of fields from the Weyl multiplet. The quick glance on the last term above indicates that the genuine N=2 supersymmetric higher derivative $R^4+...$ action with auxiliary fields taking their deformed on shell values will be found to be of the Born-Infeld with higher derivative type of action, with higher and higher powers of local graviphoton couplings $\pa^{m} F^{n}$. The basic reason for this is simple: in the minimal  N=2 superconformal model the auxiliary field from the superconformal Weyl multiplet has the classical EOM of the form
\be
{\cal T}_{\mu\nu} ^+= 4 {\cal F}_{\mu\nu}^+\, ,
\label{cl}\ee
which makes the on shell classical action dependence on the graviphoton of the Maxwell type.

 However, when the superconformal partner of the quartic Weyl tensor   $(C_{....})^4$
 which is of the form $[\pa {\cal T}]^4 $ is added to the action 
\be
S= S_0+ \lambda [\pa {\cal T}]^4 +\cdots,
\ee 
 the EOM for ${\cal T}_{\mu\nu}$ becomes {\it schematically}
\be
{\cal T}_{\mu\nu} ^{+def} =   4 {\cal F}_{\mu\nu}^+ + \lambda \, [\partial^4 {\cal T}^3]_{\mu\nu}^+ +\cdots.
\label{EOM}\ee
Here ... stands for other terms in the variation of the action (\ref{faction}) over ${\cal T}$.  None of the terms in ... cancels the term cubic in ${\cal T}$ above.
The explicit  expression for $[\partial^4 {\cal T}^3]$
 can be obtained by differentiating (\ref{CKO}).
The recursive solution of this EOM leads to infinite number of higher derivative terms with higher powers of $\cal F$
\be
{\cal T}^{def} =   {\cal F} + \lambda \, [ \partial^4  {\cal F}^3] +  \lambda^2 [ \partial^4  {\cal F}^2][ \partial^4  {\cal F}^3] +\cdots.
\label{1st}\ee
Here again we ignore numerical factors and index structures, referring to explicit expressions in \cite{Chemissany:2011yv}  where the analogous recursive equations were solved. However, we keep track of the number of derivatives and powers of graviphoton fields in these schematic equations.

The action has non-linear in graviphotons terms with higher derivatives \footnote{Analogous terms were shown in \cite{Chemissany:2011yv} to be required to restore the $U(1)$ duality broken by the higher derivative candidate counterterm $\pa^{4} F^{4}$.} 
\be
S^{def} =-{1\over 4} {\cal F}^2 + \lambda ([\pa {\cal F}])^{4}+ \lambda ^2 [\pa^{8} {\cal F}^{6}]+\cdots.
\label{Sdef}\ee
The new term $ \lambda ^2 [\pa^{8} {\cal F}^{6}]$ can be found in detailed form in \cite{Chemissany:2011yv}, see eq. (3.20) for the 
A-part of the expression, the B-part of the total supersymmetric expression can be derived explicitly using analogous method. 

However, the indication coming from just evaluating the effect of the quartic in ${\cal T}_{\mu\nu}$ term in the $R^4$ action may not give us a complete expression for the $\lambda^2$ terms in the action since there are many auxiliary fields which have to be simultaneously eliminated and therefore we need a systematic procedure here.

\subsection{Gauge-fixing and elimination of all auxiliaries}
We will work out the gauge-fixing procedure in the presence of higher order term in (\ref{faction}). We will use the non-linear vector multiplet for the second compensator, following \cite{deWit:1980tn,Mohaupt:2000mj}.
The second compensator in N=2 superconformal models is required since without it the EOM for the auxiliary field $D$ of the Weyl multiplet are inconsistent. We find that the choice of the non-linear vector multiplet works well for our deformed action (\ref{faction}) and permits a consistent gauge-fixing of our model to N=2 Poincar\'e supergravity with higher derivative terms.

The gauge-fixed action still has only terms linear in the deformation parameter $\lambda$ as given in (\ref{faction}) but it depends on many auxiliary fields $\phi_{aux}$ from the Weyl multiplet,  from the chiral compensator, and from the non-linear vector compensator, respectively\footnote{For more details, the reader may want to check the appendix.}
\be
 \phi_{aux}= \Big \{ ( b_\mu, {\cal T}_{ab}{}^{ij}\, , \chi^i\, ,   D\, ,   A_\mu \, ,  \mathcal{V}_\mu{}^i{}_j) \, , (Y_{ij}, \Omega_i) \, (V_a\, ,   M^{ij}, \lambda^i) \Big\}\, .
\ee

The auxiliary field $D$ acquires a value $D_{V}$ defined by the constraint from the non-linear vector compensator so that it is expresed via other fields
\be
 2 D_{V}\equiv    \Big ( {1\over 3}R - {\cal D}^a V_a  
+ \ft12 V^a V_a + \ft14 | M_{ij} |^2 - D^a \Phi^i_{\;\;\alpha}
D_a \Phi^{\alpha}_{\;\;i} + 2[\bar \lambda_i( \gamma^a D_a \lambda^i + {3\over 2} \chi^i -{1\over 4} \sigma\cdot {\cal T} ^{-ij} \lambda_j)+h.c.]\Big).
\label{V}\ee
The reason for using the non-linear vector multiplet as a second compensator is that the constraint (\ref{V}) is a necessary requirement for the closure of the supersymmetry algebra \cite{deWit:1980tn}. Therefore it is well suited for superconformal models deformed by higher derivative actions. 

The common feature of all these fields is that they enter in the  classical part of the superconformal action algebraically, without derivatives. The gauge-fixing of extra local symmetries allows to eliminate some of these auxiliary fields. For example special conformal K-symmetry and S-supersymmetry are broken by the gauge-fixing condition
\be
b_\mu=0\, , \qquad   \Omega_i=0,
\ee
respectively, which eliminates these two fields. The remaining auxiliary fields are eliminated by a combination of solving  EOM and gauge-fixing. For example, the $U(1)$ connection $A_\mu$ and the $SU(2)$ connection $\mathcal{V}_\mu{}^i{}_j$ are  independent auxiliary fields which enter the superconformal classical action algebraically without derivatives. However, when the classical EOM are solved, they become functions of other fields and their derivatives and as a result of gauge-fixing and solving classical EOM both
 of these connections  vanish. Thus what we call $ \phi_{aux}^0$, the values of auxiliaries in gauge-fixed classical theory without higher derivatives (only symmetries of supergravity are not fixed) are
 \bea
 &&\phi_{aux}^0= \Big \{ ( b_\mu=0, {\cal T}_{ab}{}^{(0)+}=4  {\cal F}_{\mu\nu}^+\, , \chi^i=0\, ,   D^{(0)}=D_V^{(0)}\, ,   A_\mu=0 \, ,  \mathcal{V}_\mu{}^i{}_j=0) \, , \nonumber\\
 &&(Y_{ij}=0, \Omega_i=0) \, (V_a=0\, ,   M^{ij}=0, \lambda^i=0) \Big\},
\label{cla}\eea
i.e., the auxiliary field ${\cal T}_{\mu\nu} $ is replaced by the graviphoton $F_{\mu\nu}$, and the value of $D_{V}^{(0)}$ from the second compensator removes the inconsistency of classical $D$ field equation in a model with only one compensator.
\be
 {\cal T}_{ab}{}^{(0)+}=4  {\cal F}_{\mu\nu}^+ \qquad D_{V}^{(0)} =  {1\over 6} R, 
\label{Tcl}\ee
and all other auxiliary fields vanish in classical theory without deformation.

In presence of a $\lambda$ term in (\ref{faction})  each of the classical values of  auxiliaries, except the ones which were gauge-fixed to vanish, like $b_\mu, \Omega_i $  may acquire, in principle, a non-vanishing value proportional to the first power of $\lambda$ and higher powers of this deformation parameter.
\bea
&&{\cal T}_{\mu\nu} ^+- 4  {\cal F}_{\mu\nu}^+ =  \sum_{n=1}\lambda^n {\cal T}_{\mu\nu} ^{+(n)} \, , \quad  D=D_V^{(0)}+ \sum_{n=1}\lambda^n D^{(n)}\, ,  \quad  \mathcal{V}_\mu{}^i{}_j= \sum_{n=1}\lambda^n (\mathcal{V}_\mu{}^{i }{}_j)^{(n)},\nonumber \\
&&
A_\mu = \sum_{n=1}\lambda^n A_\mu ^{(n)}\, ,\quad Y_{ij}= \sum_{n=1}\lambda^n Y_{ij} ^{(n)}\, , \quad V_a= \sum_{n=1}\lambda^n  V_a^{(n)}\, , \quad M^{ij}= \sum_{n=1} \lambda^n (M^{ij})^{(n)}.
 \label{cla1}\eea
Thus, we have to compute the expression for the N=2 supergravity action  after gauge-fixing the extra symmetries in presence of deformation by eliminating all auxiliaries on their EOM. 
For this purpose we need to have an explicit expression for the gauge-fixed action starting with the superconformal one in (\ref{faction}). 


\section{ Deformation of the super-Poincar\'e supersymmetry}

In order to study the deformation of supersymmetry along the lines explained in the introduction we need to know the $\lambda$ level of auxiliary fields which enter the classical superconformal transformation of gravitino
\begin{eqnarray}
 \delta \psi_{\m}^{i \, Superconf}&=&2 {\cal D}_{\m}\e^{i}-\frac{1}{16}\g_{ab} {\cal T}^{-\,ab }\g_{\m} \epsilon^{ij}\e_j
 -\g_{\m}\eta^{i}\label{gravitino} \, ,
\end{eqnarray}
 where the N=2 superconformal derivative of the spinor reads
 \be
{\cal D}_{\m} \e^{i}= (\pa_{\m}-\frac{1}{4}\omega_{\m}{}^{cd}\g_{cd}+\frac{1}{2}b_{\m}+\frac{1}{2}i A_{\m})\e^{i}+\frac{1}{2}\mathcal{V}_{\m}{}^{i}_{j}\e^{j},
 \ee
 and the spin connection is $\omega_{\m}{}^{cd}$, the dilatational connection is $b_\mu$, the chiral $U(1)$ connection is $A_\mu$ and the $SU(2)$ connection is $\mathcal{V}_{\m}{}^{i}_{j}$.
In notation of \cite{deWit:2010za}
 $\epsilon^i$ stands for 8 Q-supersymmetry parameters and 
  and $\eta^i$ for 8 special S-supersymmetry parameters. 
  When the S-supersymmetry is gauge-fixed one finds the dependence on the compensating S-supersymmetry parameter $\eta(\epsilon)$ which preserves the gauge
  \be
  \Omega_i=0\, .
  \ee
We now use eq. (D.3) from \cite{deWit:2010za}
\be
\delta \Omega_i= 2\gamma^\mu D_\mu X\epsilon_i + \frac{1}{2} F^{-}_{ab} \gamma^{ab}\varepsilon_{ij} \epsilon^j + Y_{ij} \epsilon ^j + 2X \eta_i =0,
\ee
where
\begin{eqnarray}
\label{FS}
 F_{ab}^-= \left(dW\right)^{-}_{ab}  
  +\tfrac14\big[\bar{\psi}_{\rho}{}^i\gamma_{ab} \gamma^\rho\Omega^{j}
  + \bar{X}\,\bar{\psi}_\rho{}^i\gamma^{\rho\sigma}\gamma_{ab}
  \psi_\sigma{}^j
  - \bar{X}\, {\cal T}_{ab}{}^{ij}\big]\varepsilon_{ij}  \,,
\end{eqnarray}
 This defines for us the local supergravity supersymmetry as a combination of the superconformal Q- and S-supersymmetry transformation:
\be
\eta_i(\epsilon)=-\left(\gamma^\mu \frac{D_\mu X}{X}\epsilon_i + \frac{1}{4 X} F^{-}_{ab} \gamma^{ab}\epsilon_{ij} \epsilon^j + \frac{Y_{ij}}{2X} \epsilon ^j \right),
\ee
We can now write for the gravitino local supersymmetry transformation in supergravity using the gauge fixing conditions 
\be
X=\bar X=1\, , \qquad b_\mu=0\, , \qquad \Phi_\alpha^i=\delta_\alpha^i,
\ee 
\begin{eqnarray}
&& \delta \psi_{\m}^{i \, SG} =2  D_{\m}^{SG} \e^{i}-\frac{1}{16}\g_{ab} {\cal T}^{-\, ab\, ij}\g_{\m}\e_{j}\nonumber\\
&& - \frac{1}{4 } F^{-}_{ab}  \g_{\m}\gamma^{ab}\e^{ij} \epsilon_{j}
  + 3i [\g_\m, \gamma_\nu]  A^{\nu} \epsilon^{i} + (\mathcal{V}_{\m}{}^{i}_{j}+\frac{1}{2} Y^{i}{}_{j}\g_{\m}) \epsilon ^j
\label{gravitino2}, \end{eqnarray}
 \be
 D_{\m} ^{SG}\e^{i}= (\pa_{\m}-\frac{1}{4}\omega_{\m}{}^{cd}\g_{cd})\e^{j}. 
 \ee
Now note that the classical $ F_{ab}$, being a component of the compensator, vanishes when EOM for ${\cal T}$ are used
\begin{eqnarray}
\label{FS1}
 \big(F^{(0)}\big)_{ab}^-= 0\, , \qquad \Rightarrow \quad \big( {\cal T}^{(0)}\big)_{ab}^{-}\equiv  4 \left(dW\right)^{-}_{ab}  
  + \bar{\psi}_\rho{}^i\gamma^{\rho\sigma}\gamma_{ab}
  \psi_\sigma{}^j
    \,,
\end{eqnarray}
If we use
\be
\phi_{aux} = \phi_{aux} ^0 + \Delta \phi_{aux},
\ee
we see that
\be
\Delta  F_{ab}^-= -{1/4} \Delta  {\cal T}_{ab}{}^{-},
\ee
\begin{eqnarray}
&& \delta \psi_{\m}^{i \, SG} =2  D_{\m}^{SG} \e^{i}-\frac{1}{16}\g_{ab} ({\cal T}^{-\, ab})^0\g_{\m}\e^{i}\nonumber\\
\nonumber\\
&&-\frac{1}{16} \Delta {\cal T}^{-\, ab}[ \g_{ab}, \g_{\m}]\e^{i}  
  + 3  i  \Delta A^{\nu} [\g_\m, \gamma_\nu] \epsilon^{i} +  \Delta (\mathcal{V}_{\m}{}^{i}_{j}+\frac{1}{2} Y^{i}{}_{j}\g_{\m}) \epsilon ^j
\label{gravitinoDelta}. \end{eqnarray}
Here the 1st line has a standard undeformed N=2 supergravity transformation of gravitino, the second line is present due to higher derivative term in the action which induces the non-vanishing $\Delta \phi_{aux}$. Each term in the second line vanishes at $\lambda=0$ and is not vanishing, a priory, for non-vanishing $\lambda$. Our goal is to study terms linear in $\lambda$ in
\be
 \Delta \phi_{aux}= \sum_{n=1} \lambda^n \phi_{aux}^{(n)},
\ee
and to show that terms with $({\cal T}^{(1)-})_{ab}$  do not cancel. Making use of $\gamma_{\mu}\gamma_{ab} = \gamma_{ab} \gamma_{\mu} + 4 g_{\mu [a} \gamma_{b]},$ up to first order in $\lambda,$ Eq.(\ref{gravitinoDelta}) becomes
\begin{eqnarray}
&& \delta \psi_{\m}^{i \, SG} =2  D_{\m}^{SG} \e^{i}-\frac{1}{16}\g_{\lambda \sigma } ({\cal T}^{-\, \lambda \sigma })^0\g_{\m}\e^{i}\nonumber\\
\nonumber\\
&&+\l \Big[\frac{1}{4} ({\cal T}^{(1)-})_{\m}{}^{\n}\g_{\n}\e^{i}  
  + 3  i  A^{(1)\n} [\g_\m, \gamma_\nu] \epsilon^{i} +   (\mathcal{V}^{(1)}{}_{\m}{}^{i}_{j}+\frac{1}{2} Y^{(1)}{}^{i}{}_{j}\g_{\m}) \epsilon ^j\Big]
\label{gravitinoDeltavalue}.\end{eqnarray}
Defining
\begin{equation}({\cal T}^{(1)-})_{\m}{}^{\n}=-16  [\pa^{4} {\cal F}^{3}]_{\m}{}^{\n}+...,\qquad {\cal F}_{\m\n}=2\pa_{[\m} W_{\n]}, \end{equation}
we find
  \begin{eqnarray}
  &&\delta \psi^{(1)}{}_{\m\, }^{i} =  -4 [ \partial^4 {\cal F}^3]_{\mu}{}^{\nu}+\cdots ] \gamma_{\nu}\,\epsilon^{i}+ \nonumber\\
  &&3  i   [\g_\m, \gamma_\nu] A^{(1)\n}\epsilon^{i}+\frac{1}{2}\,\mathcal{V}^{(1)}{}_{\m}{}^{i}_{j}\,\e^{j}+\frac{1}{2}\gamma_{\mu} Y^{(1)}{}^{i}_{j}\,\e^{j}  .
  \label{psidef}\end{eqnarray}
The first term in gravitino susy rule deformation, $-4 [ \partial^4 {\cal F}^3]_{\mu}{}^{\nu} \gamma_{\nu}\,\epsilon^{i}$ is not canceled by any other term in eq. (\ref{psidef}). The terms ... have been checked by direct inspection, the terms in the second line of eq.  (\ref{psidef}) have different matrix structure in Lorentz $\gamma$-matrices and in $SU(2)$ structures. Therefore, even without using the precise (extremely complicated) expressions for all terms in eq. (\ref{psidef}) we claim that the deformation in the cubic graviphoton sector is necessary.

This is in sharp contrast with pure N=0 gravity where
\be
\delta g_{\mu\nu} = D_\mu \xi_\nu+  D_\nu \xi_\mu,
\label{gc}\ee
independently of the fact that $R_{\mu\nu}=0$ or not, since $D_\mu$ in (\ref{gc}) does not change if  any general covariant action with higher derivatives, for example the $\lambda (R_{....})^3$ action,  is added to the classical action. The local general covariance transformation rules do not depend on $\lambda$.

Meanwhile the  gravitino local supersymmetry transformation does depend on the form of the action: for classical theory we have the $\lambda$-independent expression and for the deformed one it is (\ref{gravitinoDeltavalue}) with explicit $\lambda$-dependence, caused by elimination of auxiliary fields which classically vanish. This reminds of  the situation with duality transformations, which depend on the choice of the action, as explained in \cite {Kallosh:2011dp} with regard to $E_{7(7)}$  symmetry.
 
\section{Deformation of the action}
We describe the deformed superconformal action (\ref{faction}) in details in Appendix B. The general form of this action was derived in  \cite{deWit:2010za} and we present it for completeness in  eq. (\ref{eq:quadratic-chiral-Lagr}). For our model 
the relevant components of the superfield $\Phi=W^2 S^{-2}$ are given in eq. (\ref{C}). These in turn depend on many components fields as well as geometric curvatures, defined in terms of component fields in  \cite{deWit:2010za}. 

 The detailed version of the classical part of the action is presented in Appendix B, sec. 9.1, before and after gauge-fixing. It shows that the auxiliary fields, $A_\mu$ and ${\cal V}^i_{\mu j}$ for $U(1)$ and $SU(2)$ connection, appear in covariant derivatives from the terms $|DX |^2$ and $|D\Phi^i_\alpha|^2$. When $X=\bar X=1, \Phi^i_\alpha=\delta^i_\alpha$ are fixed, one finds that both connections vanish according to classical EOM. However, when EOM are deformed one finds that the values of these connections acquire corrections as explained in (\ref{cla1}). The fields $Y_{ij}$, $M_{ij}$, $V_a$  show up in the classical action quadratically and all of them  vanish on classical EOM. Finally, one can see from eq. (\ref{Tcl}) that ${\cal T}$ classically becomes equal to $4{\cal F}$ but in presence of deformation will have corrections as described in (\ref{EOM}) and (\ref{1st}).

If we ignore the presence of all auxiliary fields but ${\cal T}$ we have  shown that the  action consistent with the recursive solution of the deformed EOM (\ref{EOM}) will have higher and higher powers of derivatives and ${\cal F}$'s. Our procedure of elimination of 
${\cal T}$ on the deformed EOM  is related to the procedure of restoration of the $U(1)$-duality broken by the $[\pa {\cal F}]^4$ described in  \cite{Chemissany:2011yv} and also given in eq. (\ref{CKO}) here. In both cases the quartic in graviphoton terms generate all higher powers of graviphotons and derivatives, i. e.  in both cases the dependence on graviphotons proliferate in the Born Infeld with higher derivative manner. Specifically the next term in \cite{Chemissany:2011yv} is
\be
\Delta S= \lambda^2 [\pa^8 {\cal F}^6]\, ,
\ee
and an analogous one appears in N=2 supergravity from inserting the deformed value of ${\cal T}$ solving deformed EOM (\ref{EOM}) into the action (\ref{faction}).

However, the situation in N=2 supergravity is significantly more complicated than the one in \cite{Chemissany:2011yv}, which is a model with one non-linearly interacting vector field. Here we have gravitino's, a graviton, a  graviphoton and many auxiliary fields.
{\it The major question therefore is: is it possible that when all auxiliaries are eliminated on their deformed EOM, all deformation of the action with higher powers of $\lambda$ and increasing powers of $ {\cal F}$ originating from the elimination of ${\cal T}$ will cancel?}
The reason the answer is not easy to get is that the expression for the action is extremely complicated, as shown in Appendix B. To simply scan the whole action is possible, and this is one way of answering these kind of questions. However, we will employ a somewhat more systematic approach based on certain properties of the action, like that it is at least quartic in all fields of the Weyl multiplet. A direct inspection of the action is also used to confirm our conclusions.

Thus, we will focus on specific part of the model where we will be able to show that such cancellation does not take place. 
Our bosonic part of the superconformal $\lambda$-deformation of the action after gauge-fixing to Poincar\'e  depends on a conformal Weyl tensor and graviphoton, all other fields are auxiliary. Our first simplifying assumption, apart from looking only at the bosonic part of the action,  is to consider only the terms in the action with the vanishing Riemann tensor and flat vierbeins 
\be
e_\mu^a= \delta_\mu^a\, , 
 \qquad R_{\mu \nu}{}^{\lambda \delta }(e, \omega)=0\, .
\ee 
It remains to study the possible interference from elimination of other auxiliary fields from Weyl and compensator multiplets with a nonlinear term in the action $\lambda^2 [\pa^8 {\cal F}^6]$ which is due to elimination of ${\cal T}$ in presence of $\lambda [\pa^4 {\cal T}^4]$,  partner of the $\lambda R^4$ in the action.

\subsection{The role of Weyl multiplet auxiliaries}

In flat space with $e_\mu^a= \delta_\mu^a$ the quartic in $W$ action depends on ${\cal T}  $  and  other auxiliary fields from the Weyl multiplet
\be
 \phi_{aux}^W= \Big \{  {\cal T}_{ab}{}^{ij}\, ,    D\, ,   A_\mu \, ,  \mathcal{V}_\mu{}^i{}_j  \Big\}\, .
\label{auxW}\ee
We have to split $ \phi_{aux}^W$ into ${\cal T}$ and the rest, 
\be
 \phi_{aux}^{'W}= \Big \{     D\, ,   A_\mu \, ,  \mathcal{V}_\mu{}^i{}_j  \Big\}.
\label{auxW'}\ee

Each term in the $\lambda$ action depends on at least four powers of fields in (\ref{auxW}). All term quartic in ${\cal T}$ we discussed, and they are main source of the higher in $\lambda$ terms. Terms which have six powers of ${\cal T}$ in the action will not interfere with our $\lambda^2 {\cal F}^6$ terms. A direct inspection  of our action shows that terms of the form
\be
\lambda  \phi_{aux}^{'W} {\cal T}^3,
\label{danger1}\ee
are absent. Since the classical action is linear and quadratic in auxiliaries it would mean that there is a possibility to have
\be
\phi_{aux}^{'W}\sim \lambda {\cal T}^3,
\ee
and the action with all auxiliaries eliminated on deformed EOM would have additional terms from
\be
\lambda  \phi_{aux}^{'W} {\cal T}^3 \qquad \Rightarrow \lambda^2 {\cal F}^6.
\ee
Such terms would be able to interfere with analogous terms originating from elimination of ${\cal T}$ via the recursive procedure, and it would be difficult to make a conclusive statement. However, in absence of terms (\ref{danger1}) so far there is no other terms 
$\lambda^2 {\cal F}^6$. The next case to study is when there are 2 fields from  $\phi_{aux}^{'W}$ and two ${\cal T}$
\be
\lambda  (\phi_{aux}^{'W})^2 {\cal T}^2.
\label{danger2}\ee
Even if such terms are present, we find that they are not relevant: each $\phi_{aux}^{'W}$ vanishes classically, therefore each will give an additional power of $\lambda$, so these terms will be cubic in $\lambda$. Same for $\lambda  (\phi_{aux}^{'W})^3 {\cal T}$, these start with $\lambda^4$ and $\lambda  (\phi_{aux}^{'W})^4$ start with  $\lambda^5$.

Here we have to note that the auxiliary field $D$ in the classical action is absent due to the second compensator constraint, see Appendix. Any appearance of $D$ in the $\lambda$-action means that it has to be replaced by its expression via the non-linear vector multiplet
\begin{equation}
\label{Dinterms}
D = \mathcal{D}^a V_a-\frac{R}{3}-\frac{V^a V_a}{2} - \frac{\left| M\right|^2}{4} + D^a \Phi^{i}_{\alpha}D_a \Phi_{i}^{\alpha}\, .
\end{equation}
When checking the statements above about the presence/absence of certain auxiliary field dependence in the higher derivative action, we took eq. (\ref{Dinterms}) into account. Upon gauge-fixing $D$ depends on $V_a$ linearly and quadratic and on $M$ quadratic and on $\mathcal{V}_\mu{}^i{}_j $ linear and quadratic.

\subsection{The role of compensator multiplets auxiliaries}

The second compensator auxiliaries are:
\be
 \phi_{aux}^2 =\Big \{  V_a\, ,   M^{ij} \Big\}.
\ee
These fields appear in the $\lambda$-action via $D$ only. The classical action has a dependence on $V_a$ and $M$ which is quadratic in each field, therefore the value of $V_a$ or $M$ is linear in $\lambda$ and remains quadratic in $M$, so $M=0$ is a consistent solution. However, the term linear in $V_a$ from $ \mathcal{D}^a V_a$ could have produced dangerous terms upon elimination of $V_a$ on deformed EOM if there would be a term $D\, {\cal T}^3$. A direct inspection shown that such terms are absent.

The first compensator auxiliary is
\be
 \phi_{aux}^1= \Big \{  Y_{ij} \Big\}.\, 
\ee
The classical action is quadratic in $Y_{ij} $, the $\lambda$-action has terms linear and higher order in $Y_{ij} $, for example $\lambda Y_{ij}X^{ij}$. Since 
\be
Y_{ij} \sim \lambda X_{ij},
\ee
the only dangerous term would be when $ X_{ij}\sim {\cal T}^3$. This again is absent, will not cancel the $\lambda^2$ term from 
${\cal T}$-recursion, and we conclude that the term $\lambda^2 d^8 F^6$must be  present in the N=2 supersymmetric higher derivative action.

\section{Discussion}
The original goal of this work was to produce the N=2 supergravity Born-Infeld with higher derivative action, which starts with the deformation of the classical N=2 supergravity action with $R^4$ terms. This goal was achieved and we claim that the N=2 superconformal action
(\ref{faction}) upon gauge-fixing of extra local symmetries and recursive elimination of auxiliary fields, described in this paper, does  produce an action with infinitely many increasing powers of the graviphoton field strength with increasing number of derivatives. It is therefore a Born-Infeld with higher derivative N=2 supergravity action.

In the process of construction such a Born-Infelld N=2 supergravity model we have realized that the genuine N=2 supersymmetric completion of the $\lambda R^4$ super-invariant is significantly different from the on shell superspace counterterm construction. Under genuine locally supersymmetric completion of the classical action supplemented with the higher derivative $\lambda R^4+...$ super-invariant 
\be
S^{def}= S_0 +  \lambda S_{ct}\, , 
\ee
we mean a standard concept of a local symmetry where
\be
\delta_Q S^{def} = \int d^4 x {\delta S\over \delta \phi^i(x) } \delta_Q \phi^i(x)=0\, .
\label{susy}\ee
 Requirement of local symmetry means that  (\ref{susy}) must be valid off shell, when 
\be
{\delta S^{def}\over \delta \phi^i(x) }\neq 0\, .
\label{offshell}\ee
In the N=2 case it is possible to construct such an action satisfying  (\ref{susy}), (\ref{offshell}).
Despite an extremely complicated form of the superconformal action which requires various auxiliary fields from Weyl and two compensator multiplets to be present, and the gauge-fixing of extra local symmetries, we have found that the supersymmetry transformation of gravitino is affected by the presence of the higher derivative action which is a  genuine supersymmetric completion of $\lambda R^4$
\be
\delta_Q \psi_\mu^{def} = \delta_Q\psi_\mu^{(0)} + \lambda \delta_Q\psi_\mu^{(1)}+\cdots\, .
\ee
The N=2 supersymmetric action with higher derivatives when the auxiliaries are eliminated has the form
\be
S^{def}= S^{(0)} + \lambda S^{(1)}+ \lambda^2 S^{(2)}+\cdots\, .
\ee
The $\lambda$-dependent terms $\lambda \delta\psi_\mu^{(1)}$ and $\lambda^2 S^{(2)}$ which we have identified in this paper, are absent in truncation down to N=2 supersymmetry of all extended supergravity counterterms  \cite{Kallosh:1980fi,Howe:1980th,Bossard:2011tq}  based on  the   on shell superspace \cite{Brink:1979nt,Howe:1981gz} where all fields are subject to classical on shell constraint ${\delta S_0\over \delta \phi^i(x) }=0$ and all auxiliary fields are taking their classical values. Meanwhile, in genuine supersymmetric completion of the higher derivative N=2 supergravity the auxiliary fields are deformed, which leads to terms missing in all extended supergravity counterterms  \cite{Kallosh:1980fi,Howe:1980th,Bossard:2011tq}  based on  the   on shell superspace \cite{Brink:1979nt,Howe:1981gz}.

 In the past the statement was made  that the on shell counterterms violate the deformed duality symmetry current conservation \cite {Kallosh:2011dp}. One could wonder whether the prediction of the global duality symmetry  is  reliable at the quantum level. However, the requirement that the action including counterterms preserves the local supersymmetry is  a standard expectation in theories with local gauge symmetries, including local supersymmetry.  Our findings that the on shell counterterms  \cite{Kallosh:1980fi,Howe:1980th,Bossard:2011tq} violate the N=2 part of the extended local supersymmetry  may help to understand the UV finiteness properties of  perturbative extended supergravties, established in  \cite{Bern:2007hh,Bern:2008pv,Bern:2012cd,Tourkine:2012ip}.

\section{Acknowledgment}
We are grateful to J. J. M. Carrasco, S. Katmadas, A. Linde, T. Ortin, R. Roiban, A. Tseytlin and A. Van Proeyen for the stimulating discussions and to J.  Broedel  was participating at the early stages of this work. WC is supported in part by the Natural Sciences and
Engineering Research Council (NSERC) of Canada.
RK is supported by Stanford Institute for Theoretical Physics, the NSF Grant No. 0756174 and John Templeton Foundation.. The work of SF is supported by the ERC Advanced Grant no. 226455, Supersymmetry, Quantum Gravity and Gauge Fields (SUPERFIELDS) and  in part by DOE Grant DE-FG03-91ER40662. The work of CSS is supported by a CSIC JAE-predoc grant  JAEPre 2010 00613. SF, WC and CSS acknowledge the hospitality at SITP where the main part of this work was performed.

\section{Appendix A:  Setup and notation}
Supersymmetric invariants with higher-derivative couplings, based on full superspace integrals,  have been constructed explicitly 
in  N=2 supergravity in  \cite{deWit:2010za}. The new invariants are coupled to conformal supergravity and are realized off-shell. This recent work  is based on previous work on N= 2 supergravity  \cite{deRoo:1980mm},\cite{deWit:1980tn}. In particular,  in \cite{deWit:2010za} as an explicit example, many of the bosonic terms of the supergravity-coupled invariants that contain $F^4$-, $R^2F^2$-, and $R^4$-terms, are  discussed. Here $F$ denotes the abelian vector multiplet field strengths and $R$ the
Riemann tensor.

As in the past, the new supergravity actions were often produced first in the superconformal form, where they have extra gauge symmetries, in our case the symmetry is associated with the superconformal $\mathrm{SU}(2,2\vert 2)$ gauge algebra. At the later stage,  a set of symmetries which is not present in the so-called Poincare supergravity (as different from superconformal models) is gauge-fixed.  The full set of superconformal symmetries include: general-coordinate, local Lorentz, dilatation, special conformal, chiral $\mathrm{SU}(2)$ and $\mathrm{U}(1)$, supersymmetry (Q) and special supersymmetry (S) transformations. From all superconformal symmetries the models of Poincare supergravity keep only general-coordinate, local Lorentz, and  the Q supersymmetry. The local dilatation, special conformal, chiral $\mathrm{SU}(2)$ and $\mathrm{U}(1)$, and special S supersymmetry  transformations have to be gauge-fixed.  In our case such a gauge-fixing will be performed by a choice of a chiral multiplet $S$, whose first component, the scalar field $A(x)$ will be gauge-fixed to $A=1$ etc. we will  present the details below. 
 We have refrained from complete  information on our model here and refer in addition to the appendices presented in \cite{deWit:2010za}.

{\it Weyl multiplet}: The Weyl multiplet plays a central role in superconformal theory since it contains all the covariant fields and curvatures of N=2 conformal supergravity. Although it is a particular case of the anti-selfdual tensor version of the chiral multiplet with $w=1$,  that is reducible,  we will start with the short explanation of the {\it Weyl supermultiplet}. The gauge fields associated with general-coordinate transformations ($e_\mu{}^a$), dilatations ($b_\mu$), chiral symmetry ($\mathcal{V}_\mu{}^i{}_j$ and $A_\mu$) and Q-supersymmetry ($\psi_\mu{}^i$) are independent fields.  The remaining gauge fields associated with the Lorentz ($\omega_\mu{}^{ab}$), special conformal ($f_\mu{}^a$) and S-supersymmetry transformations ($\phi_\mu{}^i$) are dependent fields \cite{deWit:1980tn, deWit:1984pk, deWit:1984px}. The corresponding supercovariant curvatures and covariant fields are contained in a tensor chiral multiplet, {\it Weyl multiplet} which comprises $24+24$ off-shell degrees of freedom. 
In addition to the independent superconformal gauge fields,
it contains three other fields: a Majorana spinor doublet $\chi^i$, a scalar $D$, and a selfdual Lorentz tensor $T_{abij}$, which is anti-symmetric in $[ab]$ and $[ij]$. The Weyl and chiral weights have been collected in table \ref{table:weyl}. The spinors $\epsilon^i$ and $\eta_i$ are the positive chirality spinorial parameters associated with Q- and S-supersymmetry. The corresponding negative chirality parameters are denoted by $\epsilon_i$ and $\eta^i$. We note that hermitian conjugation is always accompanied by raising or lowering of the $\mathrm{SU}(2)$ indices.

{\it Chiral multiplet}: Chiral multiplets are complex and N=2 superspace is based on four chiral and four anti-chiral anticommuting
coordinates, $\theta^i$ and $\theta_i$, so that a scalar chiral multiplet contains two times $2^4$ field components. These multiplets
carry a Weyl weight $w$ and a chiral $\mathrm{U}(1)$ weight $c$, which is opposite to the Weyl weight, i.e. $c=-w$. The weights indicate
how the lowest-$\theta$ component of the superfield scales under Weyl and chiral $\mathrm{U}(1)$ transformations. Anti-chiral multiplets can
be obtained from chiral ones by complex conjugation, so that anti-chiral multiplets will have equal Weyl and chiral weights, hence
$w=c$. The components of a generic scalar chiral multiplet are a complex scalar $A$, a Majorana doublet spinor $\Psi_i$, a complex symmetric
scalar $B_{ij}$, an anti-selfdual tensor $F_{ab}^-$, a Majorana doublet spinor $\Lambda_i$, and a complex scalar $C$. The assignment
of their Weyl and chiral weights is shown in table~\ref{table:chiral}. We refer to the Q- and S-supersymmetry transformations for a scalar chiral multiplet of weight $w$ to  \cite{deWit:2010za}.  These transformation rules  are linear in the chiral multiplet fields, and contain also other fields associated with the conformal supergravity background, such as the self-dual tensor field $T_{abij}$ and the spinor $\chi^i$. Other conformal supergravityfields are contained in the superconformal derivatives $D_\mu$.  Products of chiral superfields constitute again a chiral superfield, whose Weyl weight is equal to the sum of the Weyl weights of the separate multiplets. Also functions of chiral superfields may describe
chiral superfields, assuming that they can be assigned a proper Weyl weight. For instance, homogeneous functions of chiral superfields of
the same Weyl weight $w$ define a chiral supermultiplet whose Weyl weight equals the product of $w$ times the degree of homogeneity.

\begin{table}[t]
\begin{tabular*}{\textwidth}{@{\extracolsep{\fill}}
    |c||cccccccc|ccc||ccc| }
\hline
 & &\multicolumn{9}{c}{Weyl multiplet} & &
 \multicolumn{2}{c}{parameters} & \\[1mm]  \hline \hline
 field & $e_\mu{}^{a}$ & $\psi_\mu{}^i$ & $b_\mu$ & $A_\mu$ &
 $\mathcal{V}_\mu{}^i{}_j$ & $T_{ab}{}^{ij} $ &
 $ \chi^i $ & $D$ & $\omega_\mu^{ab}$ & $f_\mu{}^a$ & $\phi_\mu{}^i$ &
 $\epsilon^i$ & $\eta^i$
 & \\[.5mm] \hline
$w$  & $-1$ & $-\tfrac12 $ & 0 &  0 & 0 & 1 & $\tfrac{3}{2}$ & 2 & 0 &
1 & $\tfrac12 $ & $ -\tfrac12 $  & $ \tfrac12  $ & \\[.5mm] \hline
$c$  & $0$ & $-\tfrac12 $ & 0 &  0 & 0 & $-1$ & $-\tfrac{1}{2}$ & 0 &
0 & 0 & $-\tfrac12 $ & $ -\tfrac12 $  & $ -\tfrac12  $ & \\[.5mm] \hline
 $\gamma_5$   &  & + &   &    &   &   & + &  &  &  & $-$ & $ + $  & $
 -  $ & \\ \hline
\end{tabular*}
\vskip 2mm
\renewcommand{\baselinestretch}{1}
\parbox[c]{\textwidth}{\caption{\label{table:weyl}{\footnotesize
Weyl and chiral weights ($w$ and $c$) and fermion
chirality $(\gamma_5)$ of the Weyl multiplet component fields and the
supersymmetry transformation parameters.}}}
\end{table}

{\it Vector multiplet}: Chiral multiplets of $w=1$ are  reducible. The reduced scalar chiral multiplet thus describes the covariant fields and field strength of a {\it vector multiplet}, which encompasses $8+8$ bosonic and fermionic components. Table~\ref{table:vector} summarizes the Weyl and chiral weights of the various fields belonging to the vector multiplet: a
complex scalar $X$, a Majorana doublet spinor $\Omega_i$, a vector gauge field $W_\mu$, and a triplet of auxiliary fields $Y_{ij}$. 

\begin{table}[t]
\begin{center}
\begin{tabular*}{10.8cm}{@{\extracolsep{\fill}}|c||cccccc| }
\hline
 & & \multicolumn{4}{c}{Chiral multiplet} & \\  \hline \hline
 field & $A$ & $\Psi_i$ & $B_{ij}$ & $F_{ab}^-$& $\Lambda_i$ & $C$ \\[.5mm] \hline
$w$  & $w$ & $w+\tfrac12$ & $w+1$ & $w+1$ & $w+\tfrac32$ &$w+2$
\\[.5mm] \hline
$c$  & $-w$ & $-w+\tfrac12$ & $-w+1$ & $-w+1$ & $-w+\tfrac32$ &$-w+2$
\\[.5mm] \hline
$\gamma_5$   & & $+$ & &   & $+$ & \\ \hline
\end{tabular*}
\vskip 2mm
\renewcommand{\baselinestretch}{1}
\parbox[c]{10.8cm}{\caption{\label{table:chiral}{\footnotesize
Weyl and chiral weights ($w$ and $c$) and fermion
chirality $(\gamma_5)$ of the chiral multiplet component fields.}}}
\end{center}
\end{table}

\begin{table}[t]
\begin{center}
\begin{tabular*}{6.5cm}{@{\extracolsep{\fill}}|c||cccc| }
\hline
 & & \multicolumn{2}{c}{Vector multiplet} & \\  \hline \hline
 field & $X $ & $\Omega_i$ & $W_\mu$ & $Y_{ij}$   \\[.5mm] \hline
$w$  & $1$ & $ \tfrac{3}{2} $ & 0 &  2   \\[.5mm] \hline
$c$  & $-1$ & $- \tfrac12$ & 0 &  0   \\[.5mm] \hline
$\gamma_5$   &  & $ + $ &   &       \\ \hline
\end{tabular*}
\vskip 2mm
\renewcommand{\baselinestretch}{1}
\parbox[c]{6.5cm}{\caption{\label{table:vector}{\footnotesize
Weyl and chiral weights ($w$ and $c$) and fermion
chirality $(\gamma_5)$ of the vector multiplet component fields.}}}
\end{center}
\end{table}

\section{Appendix B: Tools for N=2 Born-Infeld  Supergravity  with Higher Derivatives}

The N=2 superconformal action in (\ref{faction}) needs certain ingredients, the gauge-fixing procedure, and elimination of superconformal auxiliaries, to produce explicitly the N=2 Born-Infeld type supergravity with higher derivative terms. In absence of deformation at $\lambda=0$ this model with the choice made in (\ref{prep}) and in (\ref{H}) is a minimal pure N=2 supergravity.
Here we focus on bosonic part of the model.

$S=(A_S,0,B_{Sij},F^{-}_{S ab},0,C_{S})$ is a reduced $\omega=1$ chiral superfield that describes a vector multiplet. The relation with the fields $X,W_{\mu},Y_{ij},$ of the vector multiplet is given by\footnote{Through this section, in order to avoid any possible confusion, we name the components of a given superfield with the corresponding superfield subscript.}
\begin{equation}
\label{eq:vectormultiplet}
A_S=X,~~~B_{Sij}=Y_{ij},~~~,F^{-}_{Sab}={\cal F}^{-}_{ab}-\frac{\bar{X}}{4}T^{-}_{ab},~~~C_S=-2\Box_C \bar{X} -\frac{1}{4} F^{+}_{Sab} {\cal T}^{+ab}\, ,
\end{equation}
where $F=dW$ and 
\be
{\cal T}^{+}_{ab}={\cal T}_{ab ij} \epsilon^{ij}\, ,  \qquad {\cal T}_{ab ij} ={1\over 2}{\cal T}^{+}_{ab} \epsilon^{ij}\, ,  \qquad {\cal T}^{-}_{ab}={\cal T}_{ab}^{ ij} \epsilon_{ij}\, , \qquad {\cal T}_{ab}^{ ij} ={1\over 2}{\cal T}^{-}_{ab} \epsilon_{ij}\
\ee 

\subsection{Classical model and gauge-fixing of extra superconformal symmetries}
The classical action is given by 
\begin{equation}
S_{cl}=\int d^4\theta~ S^2+h.c.\, ,
\end{equation}
where the last expression stands for 
\begin{equation}
\label{eq:Sclassic}
S_{cl}={1\over 8}\int d^4 x e~ \left( C_{S^2} - \frac{A_{S^2}}{16} (T^{+})^2\right)+h.c.\, .
\end{equation}
The action before gauge fixing and without using a non-linear second compensator is given by
\begin{equation}
\label{eq:Sclassicinc}
S_{cl}={1\over 8} \int  d^4 x \, e\, \left\{-4 X \Box_{C} \bar{X} - \frac{X}{2} F^{+}_{S~ab} T^{+~ab} + (F^{-}_S)^2 - \frac{Y^2}{2}-\frac{X^2}{16} (T^{+})^2\right\} + h.c.\, ,
\end{equation}
where
\begin{equation}
\Box_{C} \bar{X} = \mathcal{D}^{\mu}\mathcal{D}_{\mu} \bar{X} + \bar{X}\left(\frac{R}{6}-D\right)\, ,
\end{equation}
and
\begin{equation}
{\cal F}_{ab} = (dW)_{ab} = 2\partial_{[a} W_{b]}\, ,
\end{equation}

We use now the non-linear multiplet constraint to express $D$ in terms of $V^a,~M_{ij}$ and $~\Phi^{i}_{\alpha}$, to avoid inconsistency. The bosonic part of the constraint is given by
\begin{equation}
D^{a} V_{a} -3D -\frac{V^a V_a}{2} - \frac{\left| M\right|^2}{4} + D^{a} \Phi^{i}_{~~\alpha} D_{a} \Phi^{\alpha}_{~~i} = 0\, ,
\end{equation}
and therefore we obtain
\begin{equation}
\label{eq:Dinterms}
D = \mathcal{D}^a V_a-\frac{R}{3}-\frac{V^a V_a}{2} - \frac{\left| M\right|^2}{4} + D^a \Phi^{i}_{\alpha}D_a \Phi_{i}^{\alpha}\, .
\end{equation}
Using now (\ref{eq:Dinterms}) in (\ref{eq:Sclassicinc}) we obtain
\begin{eqnarray}
S_{cl} &=& {1\over 8} \int  d^4x\, e \left\{  4  \mathcal{D}^{\mu}X\mathcal{D}_{\mu} \bar{X} -2|X|^2 R  - \frac{X}{2} F^{+}_{~ab} T^{+~ab} +  (F^{+})^2 - \frac{Y^2}{2}-\frac{X^2}{16} (T^{+})^2 + h.c\right\} \nonumber\\ &+& \int  d^4x  \, e |X|^2 \left\{\mathcal{D}^a V_a-\frac{V^a V_a}{2} - \frac{\left| M\right|^2}{4} + D^a \Phi^{i}_{\alpha}D_a \Phi_{i}^{\alpha} \right\} \, ,
\end{eqnarray}
We precisely obtain Eq. (3.111) in Ref.  \cite{Mohaupt:2000mj}, when $F=-{1\over 4 }iX^2$. The equation of motion for $T$ is given by
\begin{eqnarray}
T^{+}_{ab}= \frac{4}{X} \left(dW\right)^{+}_{ab} \, , \qquad \Rightarrow \qquad F_S^+=0.
\end{eqnarray}
Finally, gauge fixing 
\be
X=\bar X=1\, , \qquad b_\mu=0\, , \qquad \Phi_\alpha^i=\delta_\alpha^i,
\ee
we obtain
\begin{equation}
S_{cl}=\int d^4x \, e \left\{ -{1\over 2 }  R - \frac{1}{16} \left( F^{+}_{S~ab} T^{+~ab} + F^{-}_{S~ab} T^{-~ab} \right) + {1\over 8}[(F^{-}_S)^2 + (F^{+}_S)^2] -\frac{1}{16\cdot 8} \left( (T^{+})^2 +(T^{-})^2 \right)\right\}\, .
\end{equation}
Classically, $F_S^+=0$, and the action is
\begin{equation}
S_{cl}=\int d^4x \, e \left\{ -{1\over 2 }  R  -\frac{1}{8 } \left( ({\cal F}^{+})^2 +({\cal F}^{-})^2 \right)\right\}\, ,
\end{equation}

\noindent
Our choices of the second compensator and of the gauge-fixing conditions for the extra superconformal symmetries are valid in application to the classical action as well as to a deformed one. Therefore in the next subsection we describe only the superconformally invariant higher derivative action: we use the non-linear vector multiplet as a second compensator and the same gauge-fixing condition as described in the classical case, when we evaluate the deformation of all auxiliary fields and their effect on the deformation caused by the recursive procedure for ${\cal T}$.

\subsection{N=2 Superconformal  $(C_{....})^4$}

In \cite{deWit:2010za} a higher derivative term was constructed through the formula
\begin{equation}
S_{4}=\int d^4 \theta~ \Phi \mathbb{T}(\Phi')\, ,
\end{equation}
where $\Phi$ and $\Phi'$ are $\omega = 0$ chiral superfields and $\mathbb{T}(\Phi')\sim \bar{D}^4\bar{\Phi}'$.  Taking $\Phi=\Phi'$ we obtain ($w=c=0$), eq. (4.2) in \cite{deWit:2010za}
  \begin{align}
  \label{eq:quadratic-chiral-Lagr}
  e^{-1}\mathcal{L}_{4} =&\,
  4\,\mathcal{D}^2 A_{\Phi}\,\mathcal{D}^2\bar A_{\Phi}
  + 8\,\mathcal{D}^\mu A_{\Phi}\, \big[R_\mu{}^a(\omega,e) -\tfrac13
  R(\omega,e)\,e_\mu{}^a \big]\mathcal{D}_a\bar A_{\Phi} + C_{\Phi}\,\bar C_{\Phi}
  \nonumber \\[.1ex]
  &\,
   - \mathcal{D}^\mu B_{\Phi ~ ij} \,\mathcal{D}_\mu B^{ij}_{\Phi} + (\tfrac16
   R(\omega,e) +2\,D) \,
   B_{\Phi ~ ij} B^{ij}_{\Phi} \nonumber\\[.1ex]
   &\,
   - \big[\varepsilon^{ik}\,B_{\Phi~ij} \,F^{+\mu\nu}_{\Phi} \,
   R(\mathcal{V})_{\mu\nu}{}^{j}{}_{k} +\varepsilon_{ik}\,B^{ij}_{\Phi}
   \,F^{-\mu\nu}_{\Phi} R(\mathcal{V})_{\mu\nu j}{}^k \big] \nonumber\\[.1ex]
  &\,
  -8\, D\, \mathcal{D}^\mu A_{\Phi}\, \mathcal{D}_\mu\bar A_{\Phi} + \big(8\, \mathrm{i}
  R(A)_{\mu\nu} +2\, T_\mu{}^{cij}\, T_{\nu cij}\big) \mathcal{D}^\mu
  A_{\Phi} \,\mathcal{D}^\nu\bar A_{\Phi}  \nonumber\\[.1ex]
  &\,
  -\big[\varepsilon^{ij} \mathcal{D}^\mu T_{bc ij}\mathcal{D}_\mu
  A_{\Phi}\,F^{+bc}_{\Phi} + \varepsilon_{ij} \mathcal{D}^\mu T_{bc}{}^{ij} \mathcal{D}_\mu
  \bar A_{\Phi}\,F^{-bc}_{\Phi}\big] \nonumber\\[.1ex]
  &\,
  -4\big[\varepsilon^{ij} T^{\mu b}{}_{ij}\,\mathcal{D}_\mu A_{\Phi}
  \,\mathcal{D}^cF^{+}_{\Phi ~ cb} + \varepsilon_{ij} T^{\mu
    bij}\,\mathcal{D}_\mu \bar A_{\Phi} \,\mathcal{D}^cF^{-}_{\Phi ~ cb}\big]
     \nonumber\\[.1ex]
    &\, + 8\, \mathcal{D}_a F^{-ab}_{\Phi}\, \mathcal{D}^c F^+{}_{\Phi ~ cb}  + 4\,
    F^{-ac}_{\Phi}\, F^+{}_{\Phi ~ bc}\, R(\omega,e)_a{}^b
     +\tfrac1{4} T_{ab}{}^{ij} \,T_{cdij} F^{-ab}_{\Phi} F^{+cd}_{\Phi}  \,.
\end{align}
The components of the $\Phi'=\Phi=\frac{W^2}{S^2}$ (which has $\omega=0$ and $c=0$), multiplet, to be inserted into the kinetic chiral multiplet action formula (\ref{eq:quadratic-chiral-Lagr}) are
\begin{eqnarray}A_{\Phi}&=&X^{-2} (T^{-})^{2}\nonumber,\\
B_{ij}|_{\Phi}&=&-2\Big[8 X^{-2}\varepsilon_{k(i}R({\cal V})^k{}_{j)ab} \, T^{lmab}\,\varepsilon_{lm}+(T^{-})^2 X^{-3}Y_{ij}\Big]\nonumber,\\
F^{-}_{ab}|_{\Phi}&=&-16 X^{-2}\mathcal{R}(M)_{cd}{}^{ab}(T^{-})^{cd}-2(T^{-})^{2}X^{-3}F^{-\,ab}_{S}\nonumber,\\
C|_{\Phi}&=& (T_{ab}{}^{ij}\varepsilon_{ij})^2  \left(-2 X^{-3} C_{S} - \frac{3}{2} X^{-4} \left( Y_{ij} Y^{ij}-2 (F^{-}_{~S})^2\right)\right) + X^{-2} C_{W^2}\nonumber\\&-& 16 X^{-3} Y^{ij} \varepsilon_{k(i}R({\cal
    V})^k{}_{j)ab} \, T^{lmab}\,\varepsilon_{lm}+32X^{-3} {\cal R}(M)_{cd}{}^{\!ab} \,
  T^{klcd}\,\varepsilon_{kl} F^{-}_{Sab},\label{C}\end{eqnarray}
where the bosonic components of $W^2$ are given by
 \begin{align}
  \label{eq:W-squared}
  A\vert_{W^2}   =&\,(T_{ab}{}^{ij}\varepsilon_{ij})^2\,,\nonumber \\[.2ex]
  B_{ij}\vert_{W^2}  =&\, -16 \,\varepsilon_{k(i}R({\cal
    V})^k{}_{j)ab} \, T^{lmab}\,\varepsilon_{lm} \,,\nonumber\\[.2ex]
  F^{-ab}\vert_{W^2}  =&\, -16 \,{\cal R}(M)_{cd}{}^{\!ab} \,
  T^{klcd}\,\varepsilon_{kl}  \,,\nonumber\\[.2ex]
  C\vert_{W^2} =&\,  64\, {\cal R}(M)^{-cd}{}_{\!ab}\, {\cal
    R}(M)^-_{cd}{}^{\!ab}  + 32\, R({\cal V})^{-ab\,k}{}_l^{~} \,
  R({\cal V})^-_{ab}{}^{\!l}{}_k  \nonumber \\
  &\, - 32\, T^{ab\,ij} \, D_a \,D^cT_{cb\,ij} \,.
\end{align}
The multiplet corresponding to the negative second power of the chiral compensator has the following components
 \begin{eqnarray}
A_{S^{-2}} &=& X^{-2},\\
B_{S^{-2}~ij} &=& - 2 X^{-3} Y_{ij},\\
F^{-}_{~S^{-2}} &=& -2 X^{-3} F^{-}_{S},\\
C_{S^{-2}} &=& -2 X^{-3} C_{S} - \frac{3}{2} X^{-4} \left( Y_{ij} Y^{ij}-2 (F^{-}_{~S})^2\right).
\end{eqnarray}
The rule for multiplication of  chiral multiplets $W^2$ and  $S^{-2}$ are known from the N=2 superconformal tensor calculus, see for example eq. (C.1) in  \cite{deWit:2010za}. For example, the $C_\Phi$ component of the product $\Phi= W^2 S^{-2}$ is given by  
\begin{eqnarray}
C_{\Phi} = A_{W^2} C_{S^{-2}}+A_{S^{-2}} C_{W^2} - \frac{1}{2} \varepsilon^{ik}\varepsilon^{jl} B_{S^{-2}~ij} B_{W^2~kl}+ F^{-}_{~W^2 ab} F^{-~ab}_{S^{-2}},
\end{eqnarray}
which results in (\ref{C}).
Thus we have explained all ingredients of our deformed superconformal action (\ref{faction}),  (\ref{eq:quadratic-chiral-Lagr}) and we can therefore study its structure with account of gauge-fixing described above.

\bibliography{Ntwo}

\providecommand{\href}[2]{#2}\begingroup\raggedright\begin{thebibliography}{10}

\bibitem{Deser:1977nt}
S.~Deser, J.~Kay and K.~Stelle,  {\em {Renormalizability Properties of
  Supergravity}}, Phys.Rev.Lett. {\bf 38} (1977)
527.

\bibitem{Deser:1978br}
S.~Deser and J.~Kay,  {\em {Three loop counterterms for extended
  supergravity}}, Phys.Lett. {\bf B76} (1978)
400.

\bibitem{Kallosh:1980fi}
R.~Kallosh,  {\em {Counterterms in extended supergravities}}, Phys.Lett. {\bf
  B99} (1981)
122--127.

\bibitem{Howe:1981xy}
P.~S. Howe, K.~Stelle and P.~Townsend,  {\em {Superactions}}, Nucl.Phys. {\bf
  B191} (1981)
445.

\bibitem{Bern:2007hh}
Z.~Bern, J.~Carrasco, L.~J. Dixon, H.~Johansson, D.~Kosower {\em et al.},  {\em
  {Three-Loop Superfiniteness of N=8 Supergravity}}, Phys.Rev.Lett. {\bf 98}
  (2007) 161303
[\href{http://www.arXiv.org/abs/hep-th/0702112}{{\tt hep-th/0702112}}].

\bibitem{Bern:2008pv}
Z.~Bern, J.~Carrasco, L.~J. Dixon, H.~Johansson and R.~Roiban,  {\em {Manifest
  Ultraviolet Behavior for the Three-Loop Four-Point Amplitude of N=8
  Supergravity}}, Phys.Rev. {\bf D78} (2008) 105019
[\href{http://www.arXiv.org/abs/0808.4112}{{\tt 0808.4112}}].

\bibitem{Brodel:2009hu}
J.~Broedel and L.~J. Dixon,  {\em {R**4 counterterm and E(7)(7) symmetry in
  maximal supergravity}}, JHEP {\bf 1005} (2010) 003
[\href{http://www.arXiv.org/abs/0911.5704}{{\tt 0911.5704}}].

\bibitem{Elvang:2010kc}
H.~Elvang and M.~Kiermaier,  {\em {Stringy KLT relations, global symmetries,
  and $E_{7(7)}$ violation}}, JHEP {\bf 1010} (2010) 108
[\href{http://www.arXiv.org/abs/1007.4813}{{\tt 1007.4813}}].

\bibitem{Beisert:2010jx}
N.~Beisert, H.~Elvang, D.~Z. Freedman, M.~Kiermaier, A.~Morales {\em et al.},
  {\em {E7(7) constraints on counterterms in N=8 supergravity}}, Phys.Lett.
  {\bf B694} (2010) 265--271
[\href{http://www.arXiv.org/abs/1009.1643}{{\tt 1009.1643}}].

\bibitem{Kallosh:2011dp}
R.~Kallosh,  {\em {$E_{7(7)}$ Symmetry and Finiteness of N=8 Supergravity}},
  JHEP {\bf 1203} (2012) 083
[\href{http://www.arXiv.org/abs/1103.4115}{{\tt 1103.4115}}].

\bibitem{Howe:1980th}
P.~S. Howe and U.~Lindstrom,  {\em {Higher order invariants in extended
  supergravity}}, Nucl.Phys. {\bf B181} (1981)
487.

\bibitem{Brink:1979nt}
L.~Brink and P.~S. Howe,  {\em {The N=8 supergravity in superspace}},
  Phys.Lett. {\bf B88} (1979)
268.

\bibitem{Howe:1981gz}
P.~S. Howe,  {\em {Supergravity in superspace}}, Nucl.Phys. {\bf B199} (1982)
309.

\bibitem{Bossard:2011tq}
G.~Bossard, P.~Howe, K.~Stelle and P.~Vanhove,  {\em {The vanishing volume of
  D=4 superspace}}, Class.Quant.Grav. {\bf 28} (2011) 215005
[\href{http://www.arXiv.org/abs/1105.6087}{{\tt 1105.6087}}].

\bibitem{Bern:2012cd}
Z.~Bern, S.~Davies, T.~Dennen and Y.-t. Huang,  {\em {Absence of Three-Loop
  Four-Point Divergences in N=4 Supergravity}}, Phys.Rev.Lett. {\bf 108} (2012)
  201301
[\href{http://www.arXiv.org/abs/1202.3423}{{\tt 1202.3423}}].

\bibitem{Tourkine:2012ip}
P.~Tourkine and P.~Vanhove,  {\em {An $R^4$ non-renormalisation theorem in N=4
  supergravity}}, Class.Quant.Grav. {\bf 29} (2012) 115006
[\href{http://www.arXiv.org/abs/1202.3692}{{\tt 1202.3692}}].

\bibitem{Kallosh:2012ei}
R.~Kallosh,  {\em {On Absence of 3-loop Divergence in N=4 Supergravity}},
  Phys.Rev. {\bf D85} (2012) 081702
[\href{http://www.arXiv.org/abs/1202.4690}{{\tt 1202.4690}}].

\bibitem{Bossard:2011ij}
G.~Bossard and H.~Nicolai,  {\em {Counterterms vs. Dualities}}, JHEP {\bf 1108}
  (2011) 074
[\href{http://www.arXiv.org/abs/1105.1273}{{\tt 1105.1273}}].

\bibitem{Carrasco:2011jv}
J.~J.~M. Carrasco, R.~Kallosh and R.~Roiban,  {\em {Covariant procedures for
  perturbative non-linear deformation of duality-invariant theories}},
  Phys.Rev. {\bf D85} (2012) 025007
[\href{http://www.arXiv.org/abs/1108.4390}{{\tt 1108.4390}}].

\bibitem{Chemissany:2011yv}
W.~Chemissany, R.~Kallosh and T.~Ortin,  {\em {Born-Infeld with Higher
  Derivatives}}, Phys.Rev. {\bf D85} (2012) 046002
[\href{http://www.arXiv.org/abs/1112.0332}{{\tt 1112.0332}}].

\bibitem{Broedel:2012gf}
J.~Broedel, J.~J.~M. Carrasco, S.~Ferrara, R.~Kallosh and R.~Roiban,  {\em {N=2
  Supersymmetry and U(1)-Duality}},
\href{http://www.arXiv.org/abs/1202.0014}{{\tt 1202.0014}}.

\bibitem{Kallosh:2012yy}
R.~Kallosh and T.~Ortin,  {\em {New E77 invariants and amplitudes}},
\href{http://www.arXiv.org/abs/1205.4437}{{\tt 1205.4437}}.

\bibitem{Fradkin:1979cw}
E.~Fradkin and M.~A. Vasiliev,  {\em {Minimal set of auxiliary fields in SO(2)
  extended supergravity}}, Phys.Lett. {\bf B85} (1979)
47--51.

\bibitem{Moura:2002ip}
F.~Moura,  {\em {Four-dimensional 'old minimal' N=2 supersymmetrization of
  R**4}}, JHEP {\bf 0307} (2003) 057
[\href{http://www.arXiv.org/abs/hep-th/0212271}{{\tt hep-th/0212271}}].

\bibitem{Kuzenko:2008ry}
S.~M. Kuzenko and G.~Tartaglino-Mazzucchelli,  {\em {Different representations
  for the action principle in 4D N = 2 supergravity}}, JHEP {\bf 0904} (2009)
  007
[\href{http://www.arXiv.org/abs/0812.3464}{{\tt 0812.3464}}].

\bibitem{Butter:2010jm}
D.~Butter and S.~M. Kuzenko,  {\em {New higher-derivative couplings in 4D N = 2
  supergravity}}, JHEP {\bf 1103} (2011) 047
[\href{http://www.arXiv.org/abs/1012.5153}{{\tt 1012.5153}}].

\bibitem{deWit:2010za}
B.~de~Wit, S.~Katmadas and M.~van Zalk,  {\em {New supersymmetric
  higher-derivative couplings: Full N=2 superspace does not count!}}, JHEP {\bf
  1101} (2011) 007
[\href{http://www.arXiv.org/abs/1010.2150}{{\tt 1010.2150}}].

\bibitem{Freedman:2012zz}
D.~Z. Freedman and A.~Van~Proeyen,  {\em {{\it Supergravity} (Cambridge
  University Press, 2012)}},.

\bibitem{Mohaupt:2000mj}
T.~Mohaupt,  {\em {Black hole entropy, special geometry and strings}},
  Fortsch.Phys. {\bf 49} (2001) 3--161
[\href{http://www.arXiv.org/abs/hep-th/0007195}{{\tt hep-th/0007195}}].

\bibitem{deWit:1980tn}
B.~de~Wit, J.~van Holten and A.~Van~Proeyen,  {\em {Structure of N=2
  Supergravity}}, Nucl.Phys. {\bf B184} (1981)
77.

\bibitem{deRoo:1980mm}
M.~de~Roo, J.~van Holten, B.~de~Wit and A.~Van~Proeyen,  {\em {Chiral
  superfields in N=2 supergravity}}, Nucl.Phys. {\bf B173} (1980)
175.

\bibitem{deWit:1984pk}
B.~de~Wit and A.~Van~Proeyen,  {\em {Potentials and Symmetries of General
  Gauged N=2 Supergravity: Yang-Mills Models}}, Nucl.Phys. {\bf B245} (1984)
89.

\bibitem{deWit:1984px}
B.~de~Wit, P.~Lauwers and A.~Van~Proeyen,  {\em {Lagrangians of N=2
  Supergravity - Matter Systems}}, Nucl.Phys. {\bf B255} (1985)
569.

\end{thebibliography}\endgroup
 \bibliographystyle{utphysmodb}

\end{document}